\documentclass[fleqn,usenatbib]{mnras}

\usepackage{newtxtext,newtxmath}

\usepackage[T1]{fontenc}

\DeclareRobustCommand{\VAN}[3]{#2}
\let\VANthebibliography\thebibliography
\def\thebibliography{\DeclareRobustCommand{\VAN}[3]{##3}\VANthebibliography}

\usepackage{graphicx}	
\usepackage{amsmath}	
\usepackage{threeparttable} 
\usepackage{pdflscape} 
\usepackage{etoolbox, siunitx}
\newcommand{\tess}{\emph{TESS}}
\newcommand{\gaia}{\emph{Gaia}}
\newcommand{\swift}{\emph{Swift}}
\newcommand{\galex}{\emph{GALEX}}
\newcommand{\jktebop}{\textsc{jktebop}}

\newcommand{\teff}{\ensuremath{T_\mathrm{eff}}}
\newcommand{\teffpri}{\ensuremath{T_\mathrm{eff,A}}}
\newcommand{\teffsec}{\ensuremath{T_\mathrm{eff,B}}}
\newcommand{\fbol}{\ensuremath{F_\mathrm{bol}}}
\DeclareSIUnit\Msun{M_\odot} 
\DeclareSIUnit\Rsun{R_\odot}
\DeclareSIUnit\Lsun{L_\odot}

\defcitealias{2013A&A...557A.119S}{S13}
\defcitealias{2016A&A...594A..92G}{G16}

\title[Fundamental effective temperature of LL Aqr]{Fundamental effective temperature measurements for eclipsing binary stars -- VII. The solar twin in LL~Aquarii}

\author[N. J. Miller et al.]{
N. J. Miller$^{1}$\thanks{E-mail: nikki.miller@physics.uu.se}
P. F. L. Maxted,$^{2}$
A. Hahlin,$^{1,2}$
and D. Graczyk,$^{3}$
\\
$^{1}$Department of Physics and Astronomy, Uppsala University, Box 516, S-75120 Uppsala, Sweden\\
$^{2}$ Astrophysics Group, Keele University, Staffordshire ST5 5BG, UK \\
$^{3}$ Centrum Astronomiczne im. Miko\l{}aja Kopernika, Polish Academy of Sciences, Rabia\'{n}ska 8, 87-100 Toru\'{n}, Poland 
}

\date{Accepted XXX. Received YYY; in original form ZZZ}

\pubyear{\the\year{}}

\begin{document}
\label{firstpage}
\pagerange{\pageref{firstpage}--\pageref{lastpage}}
\maketitle

\begin{abstract}
The eclipsing binary LL~Aqr is a bright V=9.32, detached system consisting of two solar-type stars in an eccentric orbit (P = 20.2\,d).
The secondary component, LL~Aqr~B, was previously found to have physical and atmospheric parameters very similar to the Sun. 
Using high-precision photometry from \tess{} along with previously published orbital solutions, we obtain updated model-independent stellar radii and masses: $R_A=1.3180\pm0.0013$\,\si{\Rsun}, $R_B=0.9927\pm0.0008$\,\si{\Rsun}, $M_A=1.1947\pm0.0009$\,\si{\Msun}, and $M_B=1.0334\pm0.0006$\,\si{\Msun}. 
We derive the bolometric flux and fundamental effective temperature for each component using observed magnitudes, flux ratios from light curves in multiple bands and angular diameters derived from the radii and parallax from long baseline interferometry, measuring the following values: \teffpri$=6242\pm50$\,K, \teffsec$=5839\pm44$\,K, with an additional 9\,K systematic error from the flux scale zero-point.
We confirm that LL~Aqr displays low stellar activity by obtaining 2\,$\sigma$ upper limits on the mean surface magnetic field strengths of 78\,G and 96\,G.
Our results suggest an age of $2.67-3.01$\,Gyr, which is consistent with previous studies.
LL~Aqr now joins an ever-growing sample of well-characterised benchmark stars that can be used to test and calibrate a wide variety of methods and techniques in stellar astrophysics.
\end{abstract}

\begin{keywords}
stars: solar-type -- binaries: eclipsing -- stars: fundamental parameters -- techniques: photometric
\end{keywords}



\section{Introduction}

LL~Aqr is a well-studied, detached F + early G-type eclipsing binary (DEB) system with an orbital period of 20.178\,days.
Its variability was first noticed thanks to the \textsl{Hipparcos} mission \citep{1997A&A...323L..49P}, with a follow-up study by \citet{2004IBVS.5557....1O} adding ground-based photometry to obtain some basic orbital properties of the system; notably an orbital period and epoch, and an approximation of eclipse depth in the V-band. Despite the limited quality and quantity of the observations, narrow eclipses and hence the detached nature of the system were already apparent in the phase-folded light curve. 
Light curves in the $UBV$ photometric bands were presented by \citet{2008MNRAS.390..958I}, covering the primary eclipse and much of the secondary eclipse. The authors used these, along with new radial velocity (RV) measurements, to obtain the first estimates of physical parameters for the system. LL~Aqr~B was found to be similar to the Sun, with a mass of $1.056\pm0.051$\,\si{\Msun} ($\sim$5\% precision) and radius of $1.005\pm0.016$\,\si{\Rsun} ($\sim$1.6\%), although hotter by 400\,K.

\citet{2013A&A...557A.119S}, hereafter \citetalias{2013A&A...557A.119S}, then revisited the system with a detailed analysis of 25 higher-precision RVs measured by \citet{2013Obs...133..156G} using the CORAVEL instrument \citep{1979VA.....23..279B}, in conjunction with the existing $UBV$ data and new WASP light curves \citep{2006PASP..118.1407P}, which provided extensive coverage of all orbital phases. The derived masses and radii were now at sub-1\% precision thanks to the improved data quality. 
The author compared LL~Aqr to several stellar evolution models and found that each model under-predicted the effective temperatures determined by \citet{2008MNRAS.390..958I} using colour indices. 
\citetalias{2013A&A...557A.119S} also checked for signs of stellar activity using several indicators: by inspecting the light curves for brightness modulations due to starspots, by searching X-ray and UV photometric catalogues for evidence of enhanced chromospheric flux emission, and by using an optical $\Delta R\sim$\,60\,000 spectrum to look for chromospheric Ca H and K emission. In doing so, they found no significant evidence for activity or magnetism. 

\citet{2016A&A...594A..92G}, hereafter \citetalias{2016A&A...594A..92G}, followed up on this work with extensive high-resolution spectroscopy from the HARPS and CORALIE spectrographs \citep{2003Msngr.114...20M, 2001Msngr.105....1Q}, further improving the precision of the derived masses to <0.1\%. They used the same ground-based light curves as \citetalias{2013A&A...557A.119S}, since space-based light curves for LL~Aqr were not available at the time, and so the precision on radii between the two studies are similar. 
They additionally performed a detailed analysis of the disentangled spectra, measuring the elemental abundances for 18 species in both components and finding good agreement with solar values \citep{2009ARA&A..47..481A}. 
\citetalias{2016A&A...594A..92G} used several methods to estimate the effective temperatures, adopting the atmospheric analysis result for \teffpri{}, which was then used to calculate \teffsec{} from the temperature ratio derived using the light curves. Notably, they found both stars to be approximately 500\,K cooler than in \citet{2008MNRAS.390..958I}, a difference which they attribute to the treatment of the interstellar extinction. 
A detailed comparison with several stellar evolution models struggled to reproduce the observed parameters of both components, generally over-predicting \teff{} and metallicity by $\sim$1\,$\sigma$, but yielding an age of $2.3-2.7$\,Gyr.

In this paper, we present an analysis of the \tess{} (Transiting Exoplanet Survey Satellite) light curve of LL~Aqr, and combine the results with a recent spectroscopic and astrometric orbital solution of the system using long-baseline interferometry \citep{2023A&A...672A.119G} to obtain updated accurate and precise measurements of the masses and radii for the two components.
We then use these radii, the parallax from \citet{2023A&A...672A.119G} and flux measurements covering near-ultraviolet, visible and near-infrared wavelengths to directly measure the fundamental effective temperatures (\teff{}) using the method first outlined in \citet{2020MNRAS.497.2899M}. 
We compare our results to GARSTEC stellar evolution models \citep{2008Ap&SS.316...99W}, and prompted by a small $\sim$mmag detection of sinusoidal variation in the \tess{} light curves, below the upper limit found by \citetalias{2013A&A...557A.119S}, we check for signs of stellar activity and magnetism, taking advantage of the improved quality of data and methods now available.
LL~Aqr, especially with its solar twin component, has potential to be a valuable addition to the sample of FGK-type benchmark DEBs with precise and homogeneous determinations of physical properties presented in this series so far.

\section{Observations \& data reduction}

\subsection{\tess{} light curve}

One primary eclipse and two secondary eclipses were observed in \tess{} sector 70, between 21st September and 13th October 2023, in the 200\,s cadence. One additional secondary eclipse of LL~Aqr was observed earlier in sector 42, however the data in this sector suffers from strong scattered light and so we chose to exclude it from our analysis.
We extracted the sector 70 light curve from target pixel files generated by the `TESS-SPOC' pipeline \citep{2020RNAAS...4..201C}, which were accessed from the \textit{Mikulski Archive for Space Telescopes} (MAST) via the \textsc{lightkurve} package \citep{2018ascl.soft12013L}. We used the pipeline target aperture mask and a custom background mask, defined as the 20\% of pixels with the lowest flux. Systematics were removed using multi-scale cotrending basis vectors (CBVs), crowding corrections were applied, and poor-quality data points were removed. When masking the eclipses, we noticed a small ($\sim$1\,mmag) sinusoidal variation in the out-of-eclipse level, which a Lomb-Scargle periodogram revealed to have a period of approximately 6.3\,days. We therefore chose to de-trend the light curve by fitting a sine curve with this period to the out-of-eclipse continuum and dividing through the entire sector. An alternative approach would be to fit the amplitude of a sine curve with this period as a scaling factor during the light curve fits, but after performing some preliminary tests on the data with both approaches, we found that there was no significant effect on the quality of the fit.

\subsection{UBV light curves}

We used light curves obtained by \citet{2008MNRAS.390..958I} in the Johnson $UBV$ bands using two telescopes at the Ege University Observatory. The observations cover most of the orbital phase, except for parts of the secondary eclipse ingress and egress.

\subsection{Multi-band photometry}

\subsubsection{Catalogue photometry}

We retrieved archival photometry for our \teff{} analysis in Section~\ref{sec:teff}: $NUV$ from $GALEX$ \citep{2005ApJ...619L...1M, 2014MNRAS.438.3111C}, $u$ and $v$ from SkyMapper DR4 \citep{2024PASA...41...61O}, $G$, $BP$ and $RP$ from \gaia~DR3 \citep{2023A&A...674A...1G}, $J$, $H$ and $Ks$ magnitudes from 2MASS \citep{2006AJ....131.1163S}, and finally $W1$, $W2$ and $W3$ magnitudes from the WISE All-Sky Release Catalogue \citep{2011ApJ...735..112J, 2012wise.rept....1C}.
Each magnitude was converted onto the AB magnitude scale to allow for comparison with synthetic magnitudes \citep{2012PASP..124..140B}.

\subsubsection{UVOT magnitudes}\label{sec:uvot}

Previous work has shown that robust observational constraints in the ultraviolet result in more reliable \teff{} measurements \citep{2020MNRAS.497.2899M}. 
While many stars were observed in the $NUV$ by \galex{}, during target selection we noticed that a significant minority of DEBs do not. Motivated by this, we applied for observations of a sample of DEBs and calibration stars with the Ultraviolet/Optical Telescope \citep[UVOT; ][]{2004SPIE.5165..262R} instrument on the Neils Gehrels \swift{} observatory \citep{2004ApJ...611.1005G}. We included LL~Aqr in our sample in order to check for consistency between \galex{} and \swift{} fluxes.
Here, we present UVOT $UVM2$ observations of LL~Aqr and 6 calibration stars taken during \swift{} Cycle 18. Additional observations of GSPC~P~041-C and GSPC~P~177-D were taken from the archive to supplement these data.
The calibration stars were selected from the CALSPEC catalogue of flux standard stars with absolute calibrated composite ultraviolet and optical spectra \citep{2014PASP..126..711B}. The sample was chosen such that the spectral types were comparable to LL~Aqr and the other DEBs, and that their ultraviolet fluxes were in a suitable range for observation with UVOT. Information on the calibration stars are listed in Table~\ref{tab:calib_info}.
LL~Aqr was also observed in the $U$ band during Cycle 18, however these observations were excluded due to saturation of the detector.
We downloaded the \swift{} observations from the HEASARC archive. We used the Level II reduced image files in our analysis, which had already been processed by the standard UVOT reduction pipeline.
We measured count rates and instrumental AB magnitudes for each star using the UVOTSOURCE tool, which is part of the HEASoft software package\footnote{https://heasarc.gsfc.nasa.gov/ftools}. 
For each image, we verified that the source did not fall within a region of the detector with low sensitivity\footnote{https://swift.gsfc.nasa.gov/analysis/uvot\_digest/sss\_check.html} \citep{2010MNRAS.406.1687B}, but fortunately no sources were affected.
We placed an 5'' aperture over the target, and six 10'' apertures surrounding the target to measure the background count rate. No aperture correction was required to bring the measurements onto the standard AB magnitude system.
The zero-points for converting UVOT count rates into AB magnitudes are described in \citet{2011AIPC.1358..373B}, which is an update to the main photometric calibration paper of \citet{2008MNRAS.383..627P}.
In order to check for any zero-point offset in the $UVM2$ band, we performed synthetic photometry using CALSPEC spectra of our calibration star sample and the \swift{} photometric response functions. We compared the difference between these synthetic magnitudes and the observed magnitudes, taking the mean and standard deviation of this offset as a correction to the AB magnitude zero-point. For stars brighter than 16~mag in the $UVM2$ band, such as LL~Aqr, we find a mean zero-point offset of $-0.019\pm0.027$~mag. 

\begin{table}
    \caption{The sample of calibration stars observed in the UVOT $UVM2$ band during \swift{} Cycle 18.}
    \label{tab:calib_info}
    \begin{center}
    \begin{tabular}{lccc}
        \hline
        \noalign{\smallskip}	
        Name	&	RA (J2000)	&	Dec (J2000)	&	Spectral type	\\ 
		\noalign{\smallskip}
        \hline
        \noalign{\smallskip}	
        HD~115169	&	13 15 47.39	&	  -29 30 21.2	&	G3V	\\ 
        GSPC~P~041-C	&	14 51 57.98	&	  +71 43 17.4	&	G0V	\\ 
        GSPC~P~177-D	&	15 59 13.58	&	  +47 36 41.9	&	G0V	\\ 
        GSPC~P~330-E	&	16 31 33.81	&	  +30 08 46.4	&	G2V	\\ 
        TYC~4433-1800-1	&	18 08 34.74	&	  +69 27 28.7	&	A3V	\\ 
        TYC~4205-1677-1	&	18 12 09.57	&	  +63 29 42.3	&	A3V	\\ 
        \noalign{\smallskip}
        \hline
    \end{tabular}
    \end{center}
\end{table}

\subsection{HARPS spectroscopy}\label{sec:spectroscopy}

At the time of this study, there were 21 HARPS spectra available in the ESO archive: 17 with R$\sim$80000, observed from December 2008 to September 2014, and 4 with R$\sim$115000, observed from October to November 2023. 
In this paper, we require spectra for two purposes: estimating the interstellar reddening for LL~Aqr in Section~\ref{sec:ebv}, and our investigation into activity and surface magnetic fields in Section~\ref{sec:magnets}
For the first case, we selected only the four spectra where the interstellar sodium doublet \ion{Na}{I}\,D was unaffected by significant blending due to nearby stellar lines. 
For the second case, due to the time-dependent blending of the binary components we needed to disentangle a set of spectra from a range of orbital phases to obtain a single, stacked spectrum for each component. For best results, we need a large set of observations with the same resolving power, therefore for this we selected all 17 available spectra with R$\sim$80000. We then performed spectral disentangling following the methods outlined in \citet{2010MNRAS.407.2383F}. 

\section{Analysis \& Results}

\subsection{Light curve fitting}

\subsubsection{\tess{} light curve}

\begin{figure}
    \centering
    \includegraphics[width=\linewidth]{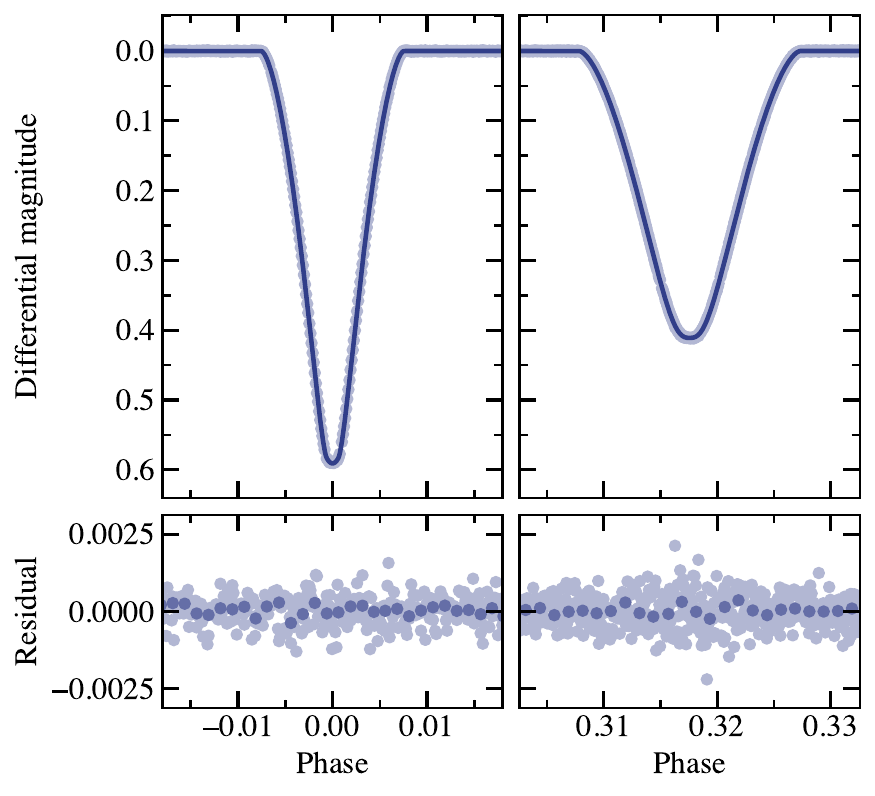}
    \caption{\tess{} sector 70 phase-folded light curve of the primary (left) and secondary (right) eclipses of LL~Aqr, along with the best-fitting \jktebop{} model following 1000 Monte-Carlo simulations. Residuals and binned residuals to the fit are given in the lower panels.}
    \label{fig:tess-lc}
\end{figure}

We fitted the \tess{} light curve using \jktebop{} version 44\footnote{https://www.astro.keele.ac.uk/jkt/codes/jktebop.html} \citepalias{2013A&A...557A.119S}, which is based on the \textsc{ebop} light curve model \citep{1981ASIC...69..111E, 1981AJ.....86..102P} and uses Levenberg-Marquardt minimisation \citep{1992nrfa.book.....P} to find optimal parameters of the model from a least-squares fit to the light curve.
We rejected 90\% of data points more than twice the eclipse duration away from each eclipse midpoint in order to speed up the fitting process.
Free parameters in the light curve fit were the surface brightness ratio ($J$), the sum and ratio of the fractional radii ($r_{sum}$ and $k$), orbital inclination ($i$), $e\cos{\omega}$, $e\sin{\omega}$, the light scaling factor, and the time of primary minimum ($T_0$). The orbital period was fixed by the ephemeris of the system, which we calculated in Section~\ref{sec:ephem}.
We varied the amount of third light ($\ell_3$), which is consistent with zero but we included it in the fit to capture the uncertainty on the remaining parameters. 
We used a quadratic limb darkening law, with initial values for the coefficients interpolated from the tables in \citet{2000A&A...363.1081C} using the \textsc{jktld} tool\footnote{https://www.astro.keele.ac.uk/jkt/codes/jktld.html}, using the $I$ band as an approximation of the \tess{} band. We included the linear coefficient for each star ($u_A$, $u_B$) as a free parameter in the fit, fixing the non-linear coefficients ($v_A$, $v_B$) at the theoretical values as setting them as free parameters has been shown to provide little improvement to the light curve fit \citep{2020MNRAS.498..332M}.
We use 1000 Monte Carlo simulations \citep{2004MNRAS.351.1277S} to obtain robust errors on the light curve fits, the results of which are given in Table~\ref{tab:lcfits}, with the best fitting model plotted alongside the data in Figure~\ref{fig:tess-lc}. We find that the secondary eclipses are total eclipses, in contrast with \citetalias{2016A&A...594A..92G}, who calculated that the secondary eclipses are partial with 99.94\% of the projected surface area obscured at mid-eclipse.

\begin{table}
\caption{Results from the \jktebop{} fit to the \tess{} sector 70 light curve.}
\label{tab:lcfits}
\begin{center}
    \begin{tabular}{@{}lccc}
        \hline
        \noalign{\smallskip}
        \multicolumn{1}{@{}l}{Parameter} & Value & Error & Note	\\
        \noalign{\smallskip}
        \hline											
        \noalign{\smallskip}											
            $T_0$	& 	$	2460223.84444	$	&	$	0.00001	$	&	BJD	\\ 
            $P$ (days)	& 	$	20.178322	$	&		---		&	fixed	\\
            $J$	& 	$	0.817	$	&	$	0.005	$	&		\\
            $r_{\rm sum}$	& 	$	0.05676	$	&	$	0.00002	$	&		\\
            $k$	& 	$	0.7532	$	&	$	0.0011	$	&		\\
            $i$ ($^{\circ}$)	& 	$	89.545	$	&	$	0.003	$	&		\\
            $e\cos{\omega}$	& 	$	-0.28823	$	&	$	0.00003	$	&		\\
            $e\sin{\omega}$	& 	$	0.1302	$	&	$	0.0007	$	&		\\
            $u_A$	& 	$	0.204	$	&	$	0.015	$	&		\\
            $u_B$	& 	$	0.247	$	&	$	0.009	$	&		\\
            $v_A$	& 	$	0.343	$	&		---		&	fixed	\\
            $v_B$	& 	$	0.311	$	&		---		&	fixed	\\
            $\ell_3$ & $-0.002$ & $0.001$ & \\
        \noalign{\smallskip}											
        \hline											
        \noalign{\smallskip}		
            Derived quantities & & & \\
            $r_A$	& 	$	0.03237	$	&	$	0.00003	$	&		\\
            $r_B$	& 	$	0.02438	$	&	$	0.00002	$	&		\\
            $e$	& 	$	0.3163	$	&	$	0.0003	$	&		\\
            $\omega$	& 	$	155.69	$	&	$	0.12	$	&		\\
            $L_B/L_A$ (\tess{})	& 	$	0.4589	$	&	$	0.0005	$	&	\\
        \noalign{\smallskip}
        \hline
    \end{tabular}
\end{center}
\end{table}

\subsubsection{UBV light curves}

\begin{figure}
    \centering
    \includegraphics[width=\linewidth]{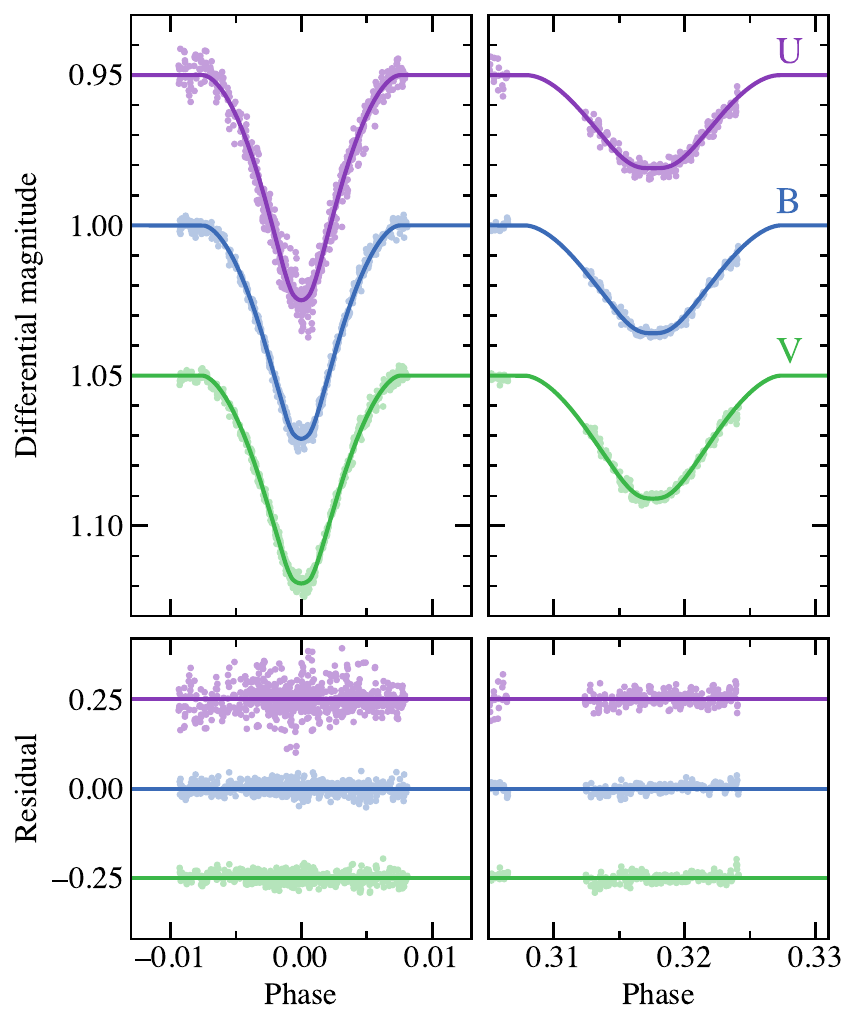}
    \caption{Phase-folded $UBV$ light curves for the primary (left) and secondary (right) eclipses, plotted alongside the best \jktebop{} fits. The residuals are shown in the lower panels, with $\pm0.25$\,mag offsets from zero for $U$ and $V$.}
    \label{fig:ubv-lc}
\end{figure}

We fitted the $UBV$ light curves with \jktebop{}, fixing the geometry of the system at the best values from our fit to the \tess{} light curves, with free parameters $J$, $T_0$, and the light scaling factor. Similarly to the \tess{} light curve fit, we used a quadratic limb darkening law for both stars, obtaining theoretical coefficients from \citet{2000A&A...363.1081C} in each photometric band, varying the linear coefficient $u_A$ for each star and fixing the non-linear coefficients. We used 1000 Monte Carlo simulations to obtain robust uncertainties on the surface brightness ratio $J$ and flux ratio $L_B/L_A$ in each band. These derived flux ratios, used in the effective temperature analysis, are given in Table~\ref{DataTable}, and the resulting light curve fits can be seen in Figure~\ref{fig:ubv-lc}.

\begin{table*}
\caption{Observed magnitudes and flux ratios for LL~Aqr, and predicted values based on synthetic photometry. The predicted magnitudes are shown with error estimates from the uncertainty on the zero-points for each photometric system. The pivot wavelength for each band is shown in the column $\lambda_{\rm pivot}$. The estimated apparent magnitudes for each star are shown given in the columns headed $m_A$ and headed $m_B$.  The flux ratio in each band is shown in the final column.}
\label{DataTable}
\centering
\begin{tabular}{@{}lrrrrrrr}
\hline
\noalign{\smallskip}
Band &  $\lambda_{\rm pivot}$ [nm]& \multicolumn{1}{c}{Observed} &\multicolumn{1}{c}{Computed} &
\multicolumn{1}{c}{$\rm O-\rm C$} &\multicolumn{1}{c}{$m_A$}  & \multicolumn{1}{c}{$m_B$}  &
\multicolumn{1}{c}{$\ell$} [\%] \\ 
\noalign{\smallskip}
\hline
\noalign{\smallskip}
UVM2 &   224.6 & $14.410\pm 0.040 $& $14.362\pm 0.026 $& $+0.048 \pm 0.048 $& $14.527\pm 0.026 $ & $16.488\pm 0.026 $ & 16.43 \\
NUV  &   230.1 & $14.212\pm 0.007 $& $14.169\pm 0.094 $& $+0.043 \pm 0.094 $& $14.362\pm 0.094 $ & $16.142\pm 0.094 $ & 19.40 \\
u    &   349.3 & $10.736\pm 0.018 $& $10.807\pm 0.030 $& $-0.071 \pm 0.035 $& $11.115\pm 0.030 $ & $12.324\pm 0.030 $ & 32.84 \\
v    &   383.6 & $10.384\pm 0.018 $& $10.406\pm 0.020 $& $-0.022 \pm 0.027 $& $10.711\pm 0.020 $ & $11.933\pm 0.020 $ & 32.46 \\
BP  &   511.0 & $ 9.378\pm 0.003 $& $ 9.380\pm 0.006 $& $-0.002 \pm 0.007 $& $ 9.752\pm 0.006 $ & $10.725\pm 0.006 $ & 40.81 \\
G    &   621.8 & $ 9.086\pm 0.003 $& $ 9.089\pm 0.008 $& $-0.003 \pm 0.008 $& $ 9.478\pm 0.008 $ & $10.390\pm 0.008 $ & 43.17 \\
RP  &   776.9 & $ 8.626\pm 0.004 $& $ 8.627\pm 0.004 $& $-0.002 \pm 0.006 $& $ 9.038\pm 0.004 $ & $ 9.882\pm 0.004 $ & 45.95 \\
J    &  1240.6 & $ 8.145\pm 0.023 $& $ 8.143\pm 0.015 $& $+0.002 \pm 0.028 $& $ 8.580\pm 0.015 $ & $ 9.341\pm 0.015 $ & 49.64 \\
H    &  1649.0 & $ 7.872\pm 0.033 $& $ 7.909\pm 0.019 $& $-0.037 \pm 0.038 $& $ 8.368\pm 0.019 $ & $ 9.066\pm 0.019 $ & 52.57 \\
Ks   &  2162.9 & $ 7.819\pm 0.023 $& $ 7.840\pm 0.030 $& $-0.021 \pm 0.038 $& $ 8.301\pm 0.030 $ & $ 8.992\pm 0.030 $ & 52.93 \\
W1   &  3389.7 & $ 7.750\pm 0.028 $& $ 7.772\pm 0.036 $& $-0.022 \pm 0.046 $& $ 8.235\pm 0.036 $ & $ 8.920\pm 0.036 $ & 53.21 \\
W2   &  4640.6 & $ 7.808\pm 0.020 $& $ 7.755\pm 0.059 $& $+0.053 \pm 0.062 $& $ 8.213\pm 0.059 $ & $ 8.912\pm 0.059 $ & 52.55 \\
W3   & 12567.5 & $ 7.802\pm 0.019 $& $ 7.788\pm 0.053 $& $+0.014 \pm 0.056 $& $ 8.242\pm 0.053 $ & $ 8.954\pm 0.053 $ & 51.91 \\
\noalign{\smallskip}
\hline
\noalign{\smallskip}
\multicolumn{5}{@{}l}{Flux ratios [\%]} \\
\noalign{\smallskip}
U    &   352.8 & $32.370 \pm 0.500 $& $32.721 $& $-0.351 \pm 0.500 $ \\
B    &   442.5 & $38.080 \pm 0.220 $& $37.671 $& $+0.409 \pm 0.220 $ \\
V    &   552.6 & $41.460 \pm 0.220 $& $42.073 $& $-0.613 \pm 0.220 $ \\
TESS &   788.0 & $45.890 \pm 0.050 $& $46.148 $& $-0.258 \pm 0.050 $ \\
K    &  2162.9 & $53.300 \pm 0.700 $& $52.750 $& $+0.550 \pm 0.700 $ \\
\noalign{\smallskip}
\hline
\end{tabular}
\end{table*}

\subsubsection{Updated orbital ephemeris}\label{sec:ephem}

We measured the time of primary minimum $T_0$ in HJD from the \tess{} light curve using \jktebop{}, fixing all parameters except $T_0$ at values from preliminary fits. We additionally obtained and fitted the WASP light curves published by \citetalias{2013A&A...557A.119S} using the same procedure, as individual eclipse times were not previously published. These times were combined with measurements from \citet{2004IBVS.5557....1O} and \citet{2008MNRAS.390..958I} to obtain the following linear ephemeris:
\begin{equation*}  
T_{\rm{pri}}(\rm{HJD/UTC}) = 2\,456\,188.17923(9) + 20.178322(1) \times E,
\end{equation*}
which is in agreement with the ephemeris published by \citetalias{2013A&A...557A.119S}.

\subsection{Masses and radii}

\citetalias{2016A&A...594A..92G} published a spectroscopic orbit using radial velocity measurements from 31 high-resolution HARPS and CORALIE \'echelle spectra with good signal-to-noise ratio ($\geq30$). \citet{2023A&A...672A.119G} later performed a simultaneous fit to these RVs and the relative positions of the two stars as measured using the GRAVITY combiner on the VLTI interferometer \citep{2011Msngr.143...16E}. 
The agreement between these two analyses is very good, therefore we used the values of $K_A$ and $K_B$ from \citet{2023A&A...672A.119G} to calculate the masses and radii of LL~Aqr, alongside $r_A$, $r_B$, $i$ and $e$ from our fit to the \tess{} light curve and $P$ from the linear ephemeris, using the equations and nominal solar constants recommended by \citet{2016AJ....152...41P} to obtain the values presented in Table~\ref{FundamentalParams}. 
Since our measurements of the stellar radii are extremely precise ($<0.2\%$), it becomes necessary for our later calculations of \teff{} to convert these photometric radii to Rosseland radii, which are used in the Stefan-Boltzmann law. This involves a small correction on the order of the scale height of the atmosphere, which we estimate by interpolating between measurements of the Sun from \citet{2009EGUGA..11.3961H} and theoretical calculations by \citet{2017AJ....154..111M}.

\begin{table}
\caption{Fundamental parameters of LL~Aqr from our analysis. Quantities are given in nominal solar units, and uncertainties on the final digit(s) are given in parentheses. For comparison, we also quote results from previous detailed studies of LL~Aqr by \citet{2016A&A...594A..92G} and \citet{2013A&A...557A.119S}.}
\label{FundamentalParams}
\begin{center}
    \begin{tabular}{@{}lcccc}
        \hline
        \noalign{\smallskip}
        \multicolumn{1}{@{}l}{Parameter} & This work & \protect{\citetalias{2016A&A...594A..92G}} & \protect{\citetalias{2013A&A...557A.119S}} \\
        \noalign{\smallskip}
        \hline
        \noalign{\smallskip}
        $M_A$ (\si{\Msun})	&	1.1947(9) & 1.1949(7) & 1.167(9)\\
        $M_B$ (\si{\Msun})	&	1.0334(6) &	1.0337(7) & 1.014(6)\\
        $R_A$ (\si{\Rsun})	&	1.3180(13) & 1.321(6) & 1.305(7)\\
        $R_B$ (\si{\Rsun})	&	0.9927(8) & 1.002(5) & 0.990(8)\\      
        $\log g_A$ (cm/s$^2$) 	&	4.2755(9) &	4.274(4) & 4.274(4)\\
        $\log g_B$ (cm/s$^2$) 	&	4.4587(8) &	4.451(4) & 4.453(7)\\
        \teffpri{} (K)	& 6242(50) & 6080(45)	& 6680(160)\\
        \teffsec{} (K)	& 5839(44) & 5703(50)	& 6200(160)	\\
        $\log L_A$ (\si{\Lsun})	& 0.376(14)	 &	0.332(14)   & 0.483(42)\\ 
        $\log L_B$ (\si{\Lsun})	& 0.013(13)	 & $-$0.019(16)	& 0.114(45)\\  
        \noalign{\smallskip}
        \hline
    \end{tabular}
\end{center}
\end{table}

\subsection{Interstellar extinction}\label{sec:ebv}

A reliable estimate of the amount of interstellar extinction is necessary for measuring accurate bolometric fluxes for stars using photometry.
Empirical and semi-empirical relations between the colour excess $E(B-V)$ and the equivalent width $EW$ of the \ion{Na}{I}\,D lines have been widely used estimating extinction. The semi-empirical relation derived by \citet{1997A+A...318..269M} based on O- and early B-type stars in the range $E(B-V)\leq 1.6$\,mag, and the empirical relation from \citet{2012MNRAS.426.1465P}, based on observations of galaxies and quasars, are two of the most commonly-used relations. Unfortunately, both suffer from a lack of data in the low-reddening regime.
\citet{maxted2025} recently addressed this issue by re-calibrating the \citet{1997A+A...318..269M} relations for both \ion{Na}{I}\,D$_1$ (5895.9\,\AA) and \ion{Na}{I}\,D$_2$ (5889.0\,\AA) in the range $E(B-V)< 0.15$\,mag. For this, they used bright B-type stars with reddening measurements from \citet{2024A&A...689A.270P} and high-resolution spectra from the IACOB spectroscopic database \citep{2015hsa8.conf..576S}.

\begin{figure}
    \centering
    \includegraphics[width=\linewidth]{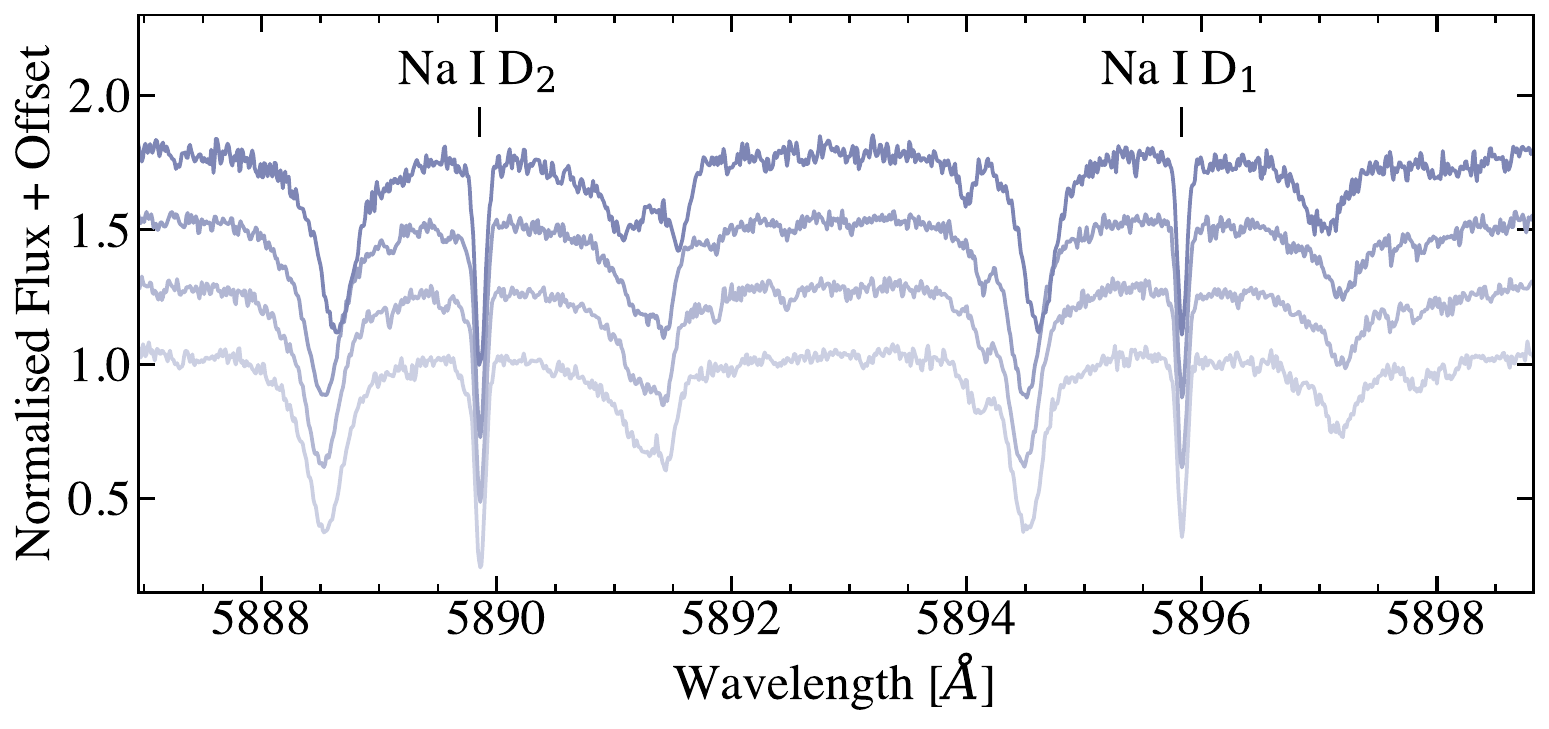}
    \caption{The HARPS spectra used in the estimation of $E(B-V)$ in the region of the \ion{Na}{I}\,D lines. The interstellar lines are clearly visible and labelled.}
    \label{fig:sodium}
\end{figure}

For LL~Aqr, we selected four HARPS spectra as described in Section ~\ref{sec:spectroscopy}. The narrow interstellar absorption lines are clearly visible in the spectra (Figure~\ref{fig:sodium}), suggesting a higher $E(B-V)$ than for stars previously analysed in this series. For each spectrum, we measured the equivalent width using numerical integration to obtain mean equivalent widths of $EW_{D1}=0.065\pm 0.001$\,\AA{} and $EW_{D2}=0.089\pm 0.001$\,\AA{}. Hence we obtain a mean reddening estimate of $E(B-V)=0.040\pm 0.015$\,mag from the \citet{maxted2025} relations, which we use as a Gaussian prior in our \teff{} analysis. 
This is higher than the value measured by \citetalias{2016A&A...594A..92G}, who obtained a mean estimate of $0.018\pm0.014$ using the \citet{2011ApJ...737..103S} extinction map and the \citet{1997A+A...318..269M} relation, so we would expect our \teff{} measurements to be slightly hotter than the photometric effective temperatures derived by \citetalias{2016A&A...594A..92G}.

\subsection{Effective temperature}\label{sec:teff}

For two stars in a detached eclipsing binary with parallax $\varpi=1/d$ and angular diameters $\theta_{A,B}=2R_{A,B}\varpi$, the total extinction-corrected flux measured at the top of the Earth's atmosphere is 
\begin{equation*}
    f_{0,b} = f_{0,A} + f_{0,B} = \frac{\sigma_{SB}}{4} \big[ \theta_A^2 T_{\rm{eff},A}^4 + \theta_B^2 T_{\rm{eff},B}^4 \big],
\end{equation*}
where $\sigma_{SB}$ is the Stefan-Boltzmann constant. All of these quantities can be measured for LL~Aqr provided that we can accurately measure the integrated fluxes $f_{0,i}$ for both stars. For this, we require observations of apparent magnitudes across the full optical range, along with measurements of the flux ratio of the binary to help separate the flux between the two stars, with the accuracy of these latter measurements being improved due to LL~Aqr displaying total eclipses.
We use the method first described in \citet{2020MNRAS.497.2899M}, which overcomes the caveats of SED fitting by using photometric data to determine the shape of the flux distribution, while using the model SEDs to provide information about the finer spectral features. To that end, we use linear superpositions of Legendre polynomials (`distortion functions') to distort the model SED for each star to find the best fit, so that the resulting best \teff{} for each star depends primarily on the data rather than choice of model. 
The distortion function for each star ($\Delta_A$ and $\Delta_B$, constructed from $N_\Delta$ polynomials with coefficients $d_{i,1}$, $d_{i,2}$ etc.) is applied to a model SED for each star to calculate synthetic photometry. The distorted SED is then normalised and integrated to calculate the total bolometric flux for each star. 
The derived effective temperatures from this method are hence relieved of the model-dependence of traditional SED fitting and based more concretely on the observed bolometric fluxes and angular diameters.
We use \textsc{emcee} \citep{2013PASP..125..306F} to sample the posterior distribution $P(M|D)\propto P(D|M)P(M)$, given the data $D$ (photometric data, observed angular diameters), and prior $P(M)$, for model parameters
\begin{equation*}
    M=\big( T_{\rm{eff},A}, T_{\rm{eff},B}, \theta_A, \theta_B, E(B-V), \sigma_{\rm m}, \sigma_{\rm r}, d_{A,1}, \dots, d_{B,1}, \dots \big)
\end{equation*}
It is unusual to have measurements of the flux ratio through the entire optical range. Such a lack of constraints may lead to unphysical or unrealistic SEDs for the two components. In order to provide some constraints on the relative fluxes of each component across all wavelengths, we assume that the colour of each star is similar to the colours of stars with the same \teff{}. We therefore set priors on the flux ratio of the binary using pre-computed empirical colour-\teff{} relations fully described in \citet{2025MNRAS.tmp.1759M}. 
The hyper-parameters $\sigma_{\rm m}$ and $\sigma_{\rm r}$ account for additional errors in the calculation of synthetic magnitudes and flux ratios, for example due to unaccounted-for uncertainties in the ZPs and photometric response functions, model SEDs, or stellar variability.
The method is implemented in the open-source software \textsc{teb}; in this work we use version v2025.08.26\footnote{https://github.com/nmiller95/teb/tree/v2025.08.26}.

\begin{figure}
    \centering
    \includegraphics[width=\linewidth]{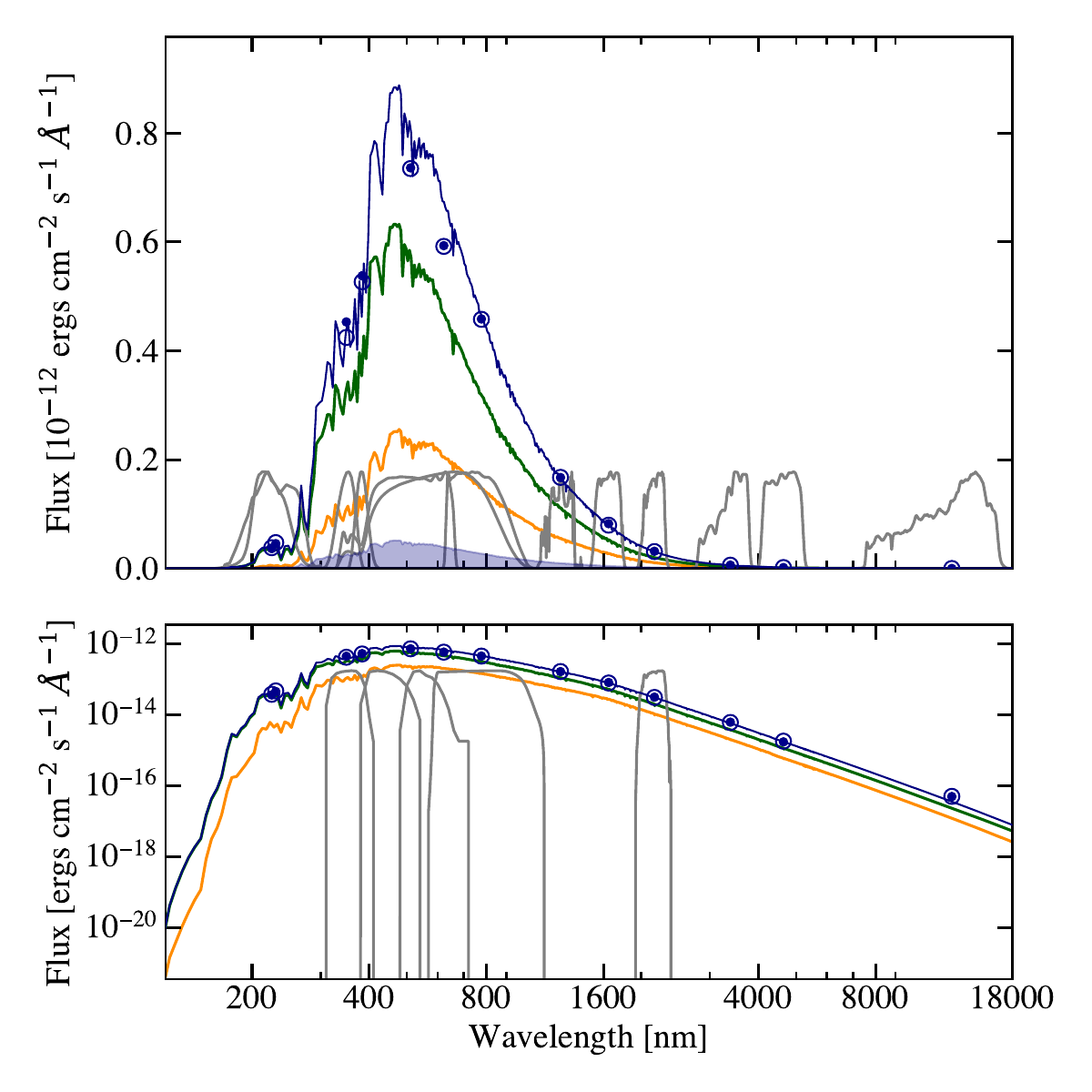}
    \caption{Upper panel: The SED of LL~Aqr. The best-fit combined SED is plotted as a line and the mean SED $\pm1-\sigma$ is plotted as a filled region. The observed fluxes are plotted as points with error bars and predicted fluxes for the best-fit SED integrated over the response functions shown are plotted with open circles. The SEDs of the two stars are also plotted with green and orange lines. Lower panel: Same as the upper panel but with fluxes plotted on a logarithmic scale. Filters used to measure flux ratios are also plotted here.}
    \label{fig:distortion}
\end{figure}

The observed photometric data we use for LL~Aqr are given in Table~\ref{DataTable}. 
For the parallax, we used the weighted average of the values measured by \citet{2023A&A...672A.119G}, $7.358\pm 0.020$\,mas, and the \gaia{}~DR3 value of $7.297\pm0.022$\,mas which includes the zero-point correction recommended by \citet{2022MNRAS.509.4276F}. 
We retrieved BT-Settl model SEDs \citep{2013MSAIS..24..128A}, calculated from PHOENIX model atmospheres, obtained from the Spanish Virtual Observatory\footnote{https://svo2.cab.inta-csic.es/theory/newov2/index.php?models=bt-settl}. We used linear interpolation between grid points to generate a model for each star at a reasonable approximation to \teff{}, $\log{g}$ and [M/H]. 
For our adopted fit, we use $T_{\rm{mod},A}=6250$\,K, $\log{g}_{\rm{mod},A}=4.28$\,dex, $T_{\rm{mod},B}=5850$\,K, $\log{g}_{\rm{mod},B}=4.46$\,dex, and [M/H]$=0.02$, [$\alpha$/Fe]$=0.0$ for both components. Previous work using \textsc{teb} has demonstrated the effects of varying model input parameters and fitting parameters in detail \citep[see][]{2020MNRAS.497.2899M, 2022MNRAS.517.5129M}, so we present only the adopted fit here. We used 256 walkers over 512 steps with an additional burn-in of 1024 steps, and four distortion polynomials for each star. The results of our adopted \teff{} fit parameters are given in Table~\ref{tab:teff-results}, with the calculated synthetic photometry and flux ratios given in Table~\ref{DataTable}. The resulting SEDs can be seen in Figure~\ref{fig:distortion}, overplotted with some of the observed photometric data. An additional systematic uncertainty on the calibration of the CALSPEC flux scale should also be taken into account \citep{2014PASP..126..711B}, which we calculated to be 9\,K for the two components. 

\begin{table}
\caption{Results from the adopted \teff{} analysis of LL~Aqr.}
\label{tab:teff-results}
\begin{center}
    \begin{tabular}{@{}lcccl}
        \hline
        \noalign{\smallskip}
        \multicolumn{1}{@{}l}{Quantity}&
        \multicolumn{1}{c}{Value}&
        \multicolumn{1}{c}{Error (stat.)}&
        \multicolumn{1}{c}{Error (sys.)}&
        \multicolumn{1}{l}{Unit}\\
        \noalign{\smallskip}
        \hline
        \noalign{\smallskip}
        \teffpri{} & 6242 & 50 & 9 & K\\
        \teffsec{} & 5839 & 44 & 9 & K \\
        \fbol$_A$ & 0.410 & 0.013 & 0.003 & $10^{-8}$\,erg\,cm$^{-2}$\,s$^{-1}$\\
        \fbol$_B$ & 0.178 & 0.005 & 0.001 & $10^{-8}$\,erg\,cm$^{-2}$\,s$^{-1}$\\
        $\theta_A$ & 0.09000 & 0.00024 & $-$ & mas \\
        $\theta_B$ & 0.06778 & 0.00017 & $-$ & mas \\
        $E$($B$-$V$) & 0.033 & 0.013 & $-$ & mag \\
        $\sigma_{\rm m}$ & 0.0090 & 0.0086 & $-$ & \\
        $\sigma_{\rm r}$ & 0.0040 & 0.0035 & $-$ & \\
        $d_{A,1}$ & $-0.003$ & 0.087 & $-$ & \\  
        $d_{A,2}$ & $-0.040$ & 0.120 & $-$ & \\ 
        $d_{A,3}$ & 0.007 & 0.069 & $-$ & \\ 
        $d_{A,4}$ & $-0.097$ & 0.094 & $-$ & \\ 
        $d_{B,1}$ & $-0.018$ & 0.086 & $-$ & \\ 
        $d_{B,2}$ & $-0.040$ & 0.140 & $-$ & \\ 
        $d_{B,3}$ & 0.009 & 0.072 & $-$ & \\ 
        $d_{B,4}$ & $-0.080$ & 0.110 & $-$ & \\ 
        \noalign{\smallskip}
        \hline
    \end{tabular}
\end{center}
\end{table}

\subsubsection{Constraints on the ultraviolet flux}

As discussed in Section~\ref{sec:uvot}, LL~Aqr is the first of a small sample of DEBs which we observed during \swift{} Cycle 18 and will analyse in future work. Unlike others in the sample, LL~Aqr has archival \galex{} observations with which we can comparatively assess the effect of including UVOT fluxes in our \teff{} analysis.
To do this, we ran \textsc{teb} with the same configuration as our adopted run, but testing different scenarios: (i) both \swift{} and \galex{}, (ii) \swift{} but no \galex{}, (iii) \galex{} but no \swift{}, and (iv) no constraints on UV flux. 
A selection of fitted parameters from these four scenarios are presented in Table~\ref{tab:teff-uv}.
We found no significant difference between the first three scenarios. Derived effective temperatures in (ii) and (iii) were at most 10 to 20\,K different to (i), so remain in agreement within error bars. Other fitted parameters $\theta_{A,B}$, $E(B-V)$ and the additional noise in magnitudes $\sigma_{\rm m}$ and flux ratios $\sigma_{\rm r}$ also varied slightly. Scenario (iii) has a smaller $\sigma_{\rm m}$ than (i) and (ii), primarily due to the smaller uncertainty in the observed $NUV$ magnitude providing a tighter constraint on the fitting procedure compared to the $UVM2$ magnitude.
Overall, the consistency of these results suggests that a single flux measurement in the $UVM2$ band, when properly calibrated to the CALSPEC flux scale, provides comparable constraints on the ultraviolet flux as the \galex{} $NUV$ band. This is an encouraging result, as it demonstrates the possibility to measure robust \teff{} for DEBs which do not currently have any catalogue ultraviolet photometry.
Scenario (iv) resulted in effective temperatures approximately 25 to 30\,K cooler than the other scenarios, and a fitted $E(B-V)$ furthest from our measured value of 0.040. This scenario also resulted in the lowest maximum log-likelihood.

\begin{table}
\caption{Selected fitted parameters from four runs of \textsc{teb} with different observational constraints in the ultraviolet. \teff{} is measured in K; $\theta$ in mas. Uncertainties on the final two digits are given in parentheses.}
\label{tab:teff-uv}
\begin{center}
    \begin{tabular}{@{}lcccc}
        \hline
        \noalign{\smallskip}
        \multicolumn{1}{@{}l}{Quantity}&
        \multicolumn{1}{c}{UVM2+NUV}&
        \multicolumn{1}{c}{UVM2 only}&
        \multicolumn{1}{c}{NUV only}&
        \multicolumn{1}{c}{No UV}\\
        \multicolumn{1}{@{}l}{}&
        \multicolumn{1}{c}{(i)}&
        \multicolumn{1}{c}{(ii)}&
        \multicolumn{1}{c}{(iii)}&
        \multicolumn{1}{c}{(iv)}\\
        \noalign{\smallskip}
        \hline
        \noalign{\smallskip}
        \teffpri{} & 6242(50) & 6255(49) & 6222(48) & 6215(49) \\
        \teffsec{} & 5839(44) & 5848(42) & 5820(43) & 5812(42) \\
        $\theta_A$ & 0.09000(24) & 0.09003(24) & 0.09002(24) & 0.09000(23) \\
        $\theta_B$ & 0.06778(17) & 0.06780(17) & 0.06780(17) & 0.06779(17) \\
        $E$($B$-$V$) & 0.033(13) & 0.037(12) & 0.027(12) & 0.024(12) \\
        $\sigma_{\rm m}$ & 0.0090(86) & 0.0082(86) & 0.0026(29) & 0.0032(41)\\
        $\sigma_{\rm r}$ & 0.0040(35) & 0.0027(25) & 0.0068(60) & 0.0062(43) \\
        $\log{\mathcal{L}}$ & 69.17 & 67.80 & 68.48 & 67.28 \\
        \noalign{\smallskip}
        \hline
    \end{tabular}
\end{center}
\end{table}

\section{Discussion}

\subsection{Comparison with stellar evolution models}

\begin{figure}
    \centering
    \includegraphics[width=\linewidth]{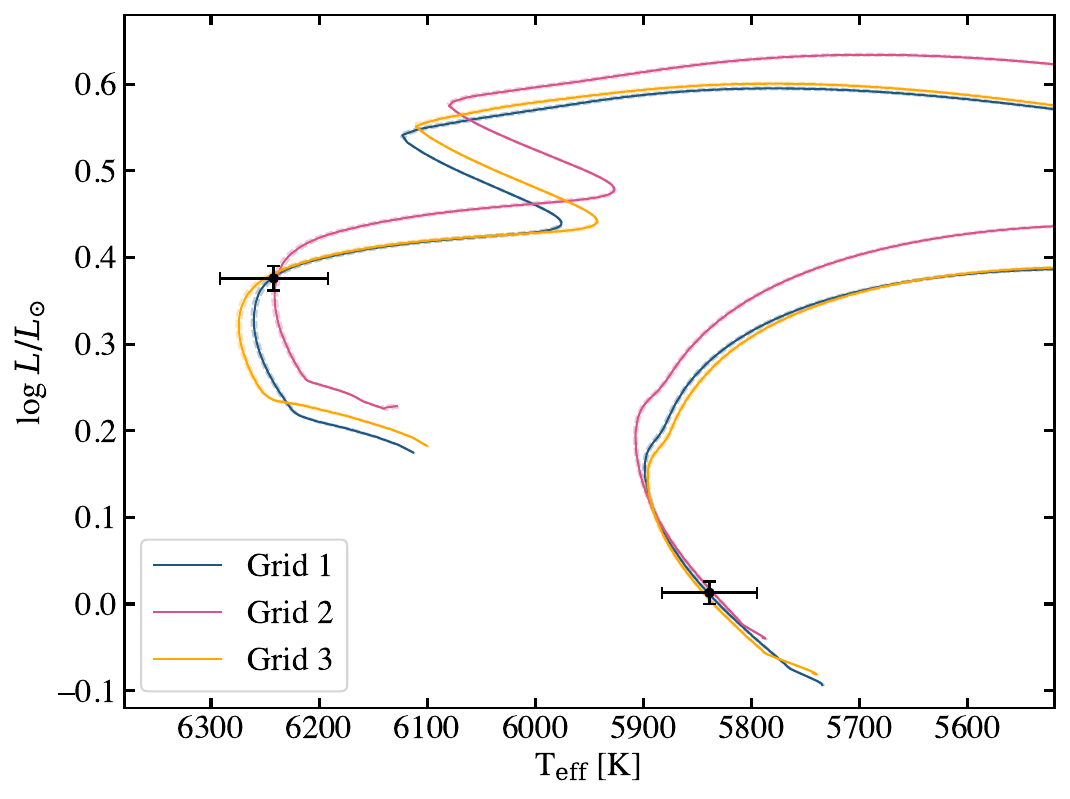}
    \caption{Hertzsprung-Russell diagram showing best-fitting GARSTEC stellar evolution tracks for all three model grids available in \textsc{bagemass} for LL~Aqr, along with observed quantities.}
    \label{fig:evotracks}
\end{figure}

As a well-detached eclipsing binary with precise physical parameters, LL~Aqr is a useful system for testing stellar evolution models. The two stars have different masses and internal structures \citepalias[see][for a detailed discussion]{2016A&A...594A..92G}, but being components in a binary, we would expect both to have formed at a common time and initial composition.
We compared the measured properties of LL~Aqr to stellar evolution tracks computed with Garching Stellar Evolution Code \citep[GARSTEC; ][]{2008Ap&SS.316...99W} using the \textsc{bagemass} code \citep{2015A&A...575A..36M}. 
This is a Bayesian Markov-chain Monte Carlo algorithm that was designed to estimate the properties of planet host stars based on their observed stellar density, \teff{}, luminosity and surface metal abundance, [Fe/H]$_{s}$. The stellar models include gravitational settling of elements so [Fe/H]$_{s}$ is typically lower than the initial metal abundance, [Fe/H]$_{i}$. Three grids of models are available within \textsc{bagemass}: one with mixing length and helium abundance calibrated on the properties of the Sun, one with a mixing length of slightly lower than the solar value ($\alpha_{\rm MLT} =1.50$ cf. $\alpha_{\rm MLT} =1.78$), and a third grid with slightly enhanced initial helium mass fraction ($Y = Y_{\odot}+\frac{dY}{dZ} + 0.02$).
We ran all three grids on both stars independently, taking values from Table~\ref{FundamentalParams} and including the mass measurements from this table as a prior in the calculation of the posterior probability distribution. We use the measurement of [Fe/H]$_{s}$ from \citetalias{2016A&A...594A..92G} but assume an increased standard error on this value of 0.1\,dex to account for our measurement of \teff{} being $\approx 150$\,K higher than the values used in their abundance analysis. The best-fitting stellar evolution tracks for each grid are shown in Figure~\ref{fig:evotracks} along with our observed quantities.

We find that first model grid with solar-calibrated values of mixing length and helium abundance gives the best individual fit to each star, and the closest agreement between the derived ages and initial metal abundance for both components. 
For the primary, we obtain a mean age of $3.01\pm0.12$\,Gyr, corresponding to an initial metal abundance [Fe/H]$_{i,A}$ of $0.075\pm0.030$. 
For the secondary component, we obtain a mean age of $2.67\pm0.12$\,Gyr, and [Fe/H]$_{i,B}=0.052\pm0.027$. Since the LL~Aqr~B is very similar to the Sun, it is interesting to note that the grid with solar values provides the best agreement to our observations. 
The third model grid with solar-calibrated mixing length and enhanced helium abundance also gives a satisfactory fit to the observations, with ages of $2.63\pm0.11$\,Gyr and $2.17\pm0.12$\,Gyr for the primary and secondary components respectively, and a higher [Fe/H]$_{i}$ compared to the first grid, at $0.156\pm0.030$ and $0.142\pm0.029$\,dex.
The second model grid with a lower mixing length value gives the largest discrepancy between the ages of the two components: $2.20\pm0.08$\,Gyr and $1.06\pm0.08$\,Gyr. The initial metal abundance is much lower than the other two grids, at $-0.006\pm0.028$ and $-0.068\pm0.026$\,dex.
Unlike \citetalias{2016A&A...594A..92G}, we do not find that the models are too hot to provide a reasonable fit to the observed properties of LL~Aqr. This discrepancy is primarily solved due to our measured \teff{} values being hotter. However, all three grids show some disagreement between the ages of the two stars, with the primary component systematically appearing older.
A full and detailed exploration of the parameter space for stellar models is beyond the scope of this study, but is certainly worthwhile to investigate whether any of the currently available stellar models are able to match the properties of both stars at the same age. In the future, a detailed differential abundance analysis with a high quality disentangled spectrum would provide a tighter constraint on the initial metal abundance and provide even more challenging tests of the models.

\subsection{Evidence of stellar activity}

\subsubsection{Magnetic fields}\label{sec:magnets}
Prior investigations of the stellar activity of LL~Aqr performed by \citetalias{2013A&A...557A.119S} revealed no sign of activity in either X-rays, FUV, or Ca H\&K emission. While this indicates that LL~Aqr is a weakly active star, the stellar activity was not quantified with an upper limit. To more easily facilitate the comparison of the activity level with other stars, we aim to expand the investigation of the stellar activity of LL~Aqr by directly measuring the magnetic field strengths of the two components. This parameter is useful to quantify, even as an upper limit, since an ideal solar twin should also exhibit similar activity behaviour to the Sun.

Our analysis is done by utilising the splitting of atomic lines caused by the Zeeman effect in the presence of a magnetic field. The Zeeman splitting causes both broadening and an increase in equivalent width refereed to as intensification. While the broadening caused by the Zeeman effect is primarily observed in the NIR due to the strong wavelength dependence of Zeeman splitting, intensification can be observed at optical wavelengths as the Zeeman splitting no longer needs to overcome other sources of line broadening to produce a signal. For Sun-like stars, \citet{kochukhov:2020a} identified a set of \ion{Fe}{I} lines at $\sim5500$\,\AA{} with a strong sensitivity to Zeeman intensification. This also includes the magnetically insensitive \ion{Fe}{I} 5434.52\,\AA{} line. 

We used MCMC sampling with the SoBAT library \citep{anfinogentov:2021} following the methods described in \citet{hahlin:2022} and \citet{hahlin:2023} to obtain the average surface magnetic field on the two components. This was done simultaneously for both components using the disentangled spectra from Section~\ref{sec:spectroscopy}. The synthetic spectra was generated with the polarised radiative transfer code \textsc{Synmast} \citep{Kochukhov:2007}, using MARCS model atmospheres \citep{2008A&A...486..951G}, a line list from VALD \citep{ryabchikova:2015}, and non-local thermodynamic equilibrium (NLTE) departure coefficients of \ion{Fe}{I} from \citet{amarsi:2022}. The \teff{} and $\log\,g$ are from this work as shown in Table~\ref{FundamentalParams}. The $v\sin{i}$ values of the two components were taken from \citetalias{2016A&A...594A..92G} and are 3.5 and 3.6\,km\,s$^{-1}$, respectively. The grid used in the analysis was made with bilinear interpolation in \teff{} and $\log\,g$ of the synthetic spectra generated with \textsc{Synmast}.

We used a two-component model for the magnetic field, where one part of the star, covering a surface fraction $f_B$, is magnetically active with a field strength $B$. Both $B$ and $f_B$ are free parameters and the average surface magnetic field strength is given by $\langle B\rangle=B\cdot f_B$. To account for non-magnetic effects, we also include abundance, luminosity ratio, macroturbulence, and radial velocity as free parameters. Additionally, we allow errors to be a free parameter to better account for systematic and model errors that typically dominates high resolution and S/N analysis of stellar spectra. \citet{hahlin:2023} have shown that this choice has no effect on the inference parameters besides a slight increase in the obtained uncertainties. The inference is ran until 1000 independent samples were collected calculated from the autocorrelation time of the MCMC sampling.

We find that the neither component show any significant magnetic field strengths, with 95\% upper limits of 78 and 96\,G for the primary and secondary component respectively. When using stellar parameters from \citetalias{2016A&A...594A..92G}, we obtained a non-detection on both components consistent with the results from this work. We also note that when performing the spectrum synthesis without NLTE departure coefficients we did in fact get a significant magnetic field detection on the primary corresponding to about 200\,G. This shows that accounting for NLTE effects is important in this context, as the change in line strength between LTE and NLTE can give rise to spurious magnetic field signals.

\begin{figure}
    \centering
    \includegraphics[width=\linewidth]{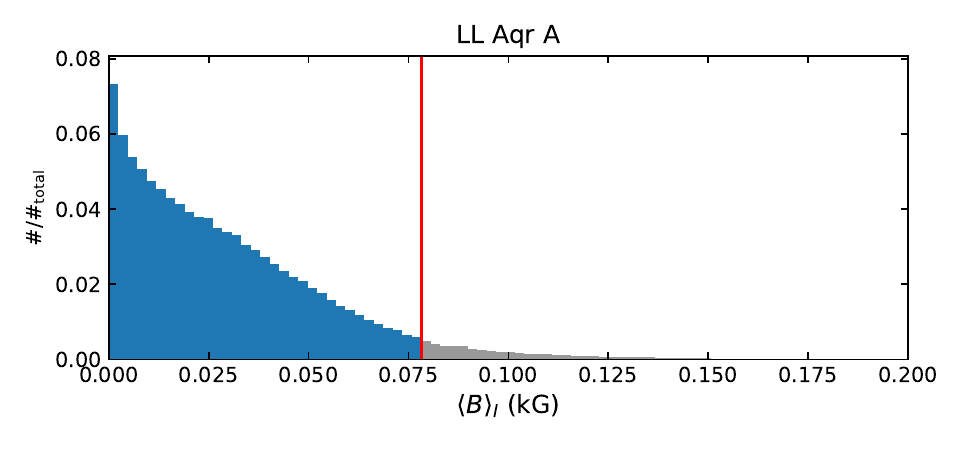}
    \includegraphics[width=\linewidth]{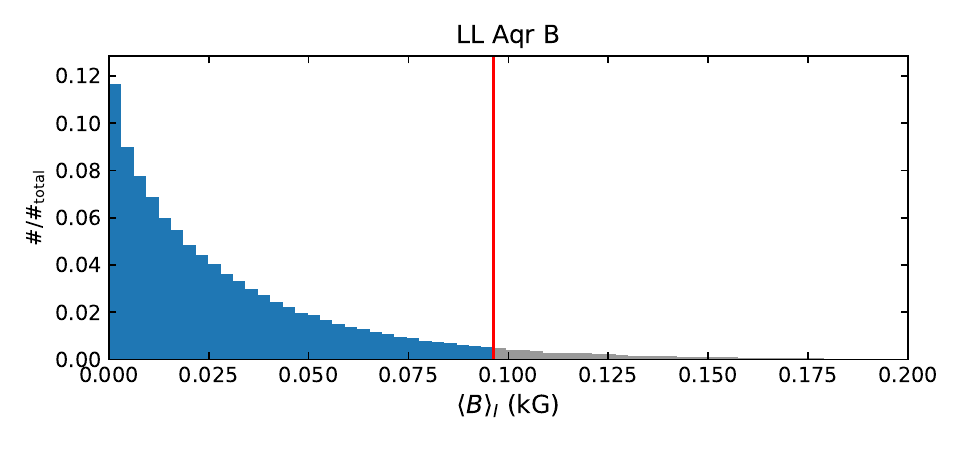}
    \caption{Posterior distribution of the average surface magnetic field strength of the two components. The red line marks the 95\,\% upper limits for the field strength.}
    \label{fig:Bupperlimits}
\end{figure}

\subsubsection{LL~Aqr in context}

When evaluating the magnetic activity of a star, it is common to use the Rossby number defined as the ratio between the rotational period and convective turnover time. While neither are given in previous works, we can estimate the two parameters from other measurements. We estimate the rotational periods by using the $v\sin{i}$ from \citetalias{2016A&A...594A..92G} and the radii obtained in this work, resulting in $P_A=19.1\pm3.2$ and $P_B=14.0\pm1.7$ days for the primary and secondary component respectively. The convective turnover times are determined using the empirical relationship from \citet{2011ApJ...743...48W} with the masses obtained from this work. This relationship gives convective turnover times of $\log\tau_A=1.04$, $\log\tau_B=1.14$, respectively. From these two parameters we obtain Rossby numbers of $1.73$ and $1.02$.

\begin{figure}
    \centering
    \includegraphics[width=\linewidth]{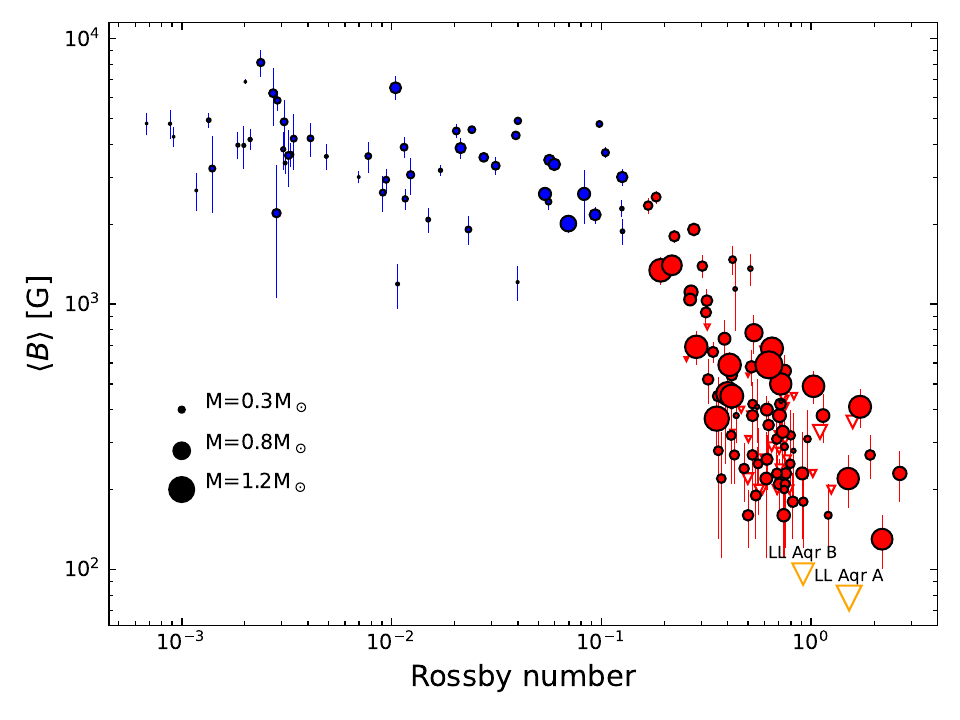}
    \caption{Magnetic field data from \citet{reiners:2022}. Blue and red points represents the separation between the saturated and unsaturated regime at a Rossby number of 0.13. Size of the symbol indicates stellar mass as shown in the figure. Upper limits for the magnetic field on LL~Aqr are the orange triangles in the bottom right.}
    \label{fig:saturation}
\end{figure}

To compare with other stars, we use the sample from \cite{reiners:2022}. While mostly focused on M dwarfs, the sample also has several sun-like stars that appear to follow the same trend between the magnetic field and Rossby number. This relationship is shown in Fig.~\ref{fig:saturation}. What we can see from this is that the obtained upper limits are consistent with magnetic field strengths one could expect at these Rossby numbers. We can also say that, even in the most magnetically active scenario, the components of LL Aqr are relatively weakly active compared to stars of similar Rossby number. Even so, an upper limit of $\sim100$\,G is still significantly stronger than the typical average field strengths on the solar surface at $\sim10$\,G \citep[e.g.][]{2023MNRAS.522.5862L}, as measured by SDO \citep{2012SoPh..275..207S}.

\section{Conclusion}

We have obtained extremely precise and accurate determinations of the masses, radii and fundamental effective temperatures for both components of LL~Aqr. 
The primary component is a mid-F type dwarf with $M_A=1.1947\pm0.0009$\,\si{\Msun}, $R_A=1.3180\pm0.0013$\,\si{\Rsun} and \teffpri{}$=6242$\,K. 
The secondary component is an early-G type dwarf with physical parameters similar to the Sun: $M_B=1.0334\pm0.0006$\,\si{\Msun}, $R_B=0.9927\pm0.0008$\,\si{\Rsun} and \teffsec{}$=5839$\,K. 
Our measured masses are in good agreement with \citet{2016A&A...594A..92G} and \citet{2023A&A...672A.119G}, while our radii are slightly smaller and significantly more precise (<0.1\% vs. <0.5\% in \citetalias{2016A&A...594A..92G}), owing to the improved quality of the \tess{} light curve.
The fundamental effective temperatures for both components, measured directly using the method first described in \citet{2020MNRAS.497.2899M}, are approximately 100-200\,K hotter than \citetalias{2016A&A...594A..92G}, but 300-400\,K cooler than \citetalias{2013A&A...557A.119S}. These large discrepancies can largely be accounted for by the different approaches each study took to estimate the interstellar reddening. We suggest that previous calibrations by e.g. \citet{1997A+A...318..269M} may have underestimated $E(B-V)$ for stars with measurable \ion{Na}{I}\,D interstellar lines at the low end of the relation due to a lack of calibrating stars with $E(B-V)<0.15$\,mag, a problem which has recently been addressed in \citet{maxted2025}.
Similarly to both \citetalias{2013A&A...557A.119S} and \citetalias{2016A&A...594A..92G}, we find that the components of LL~Aqr are approximately $1-2$\,Gyr younger than the Sun, with our best-fitting GARSTEC models predicting an age for the system of between $2.67-3.01$\,Gyr.
We placed 2\,$\sigma$ upper limits of 78\,G and 96\,G on the average surface strength of magnetic fields for each component by analysing Zeeman splitting in disentangled HARPS spectra. While these results are still $8-10$ times higher than the typical solar surface value \citep[e.g.][]{2023MNRAS.522.5862L}, we do not see significant evidence of enhanced magnetic activity in the context of Rossby number. Our results are tentatively consistent with previous investigations into activity indicators for this system presented in \citetalias{2013A&A...557A.119S}. This supports the suitability of LL~Aqr~B as a candidate solar twin, as a system with enhanced magnetic activity may not be such a useful analogue for comparative studies.
LL~Aqr is a detached, well-characterised system containing a star with similar physical and atmospheric properties to the Sun, making it a valuable addition to the growing sample of benchmark stars with direct \teff{} measurements analysed in this series.

\section*{Acknowledgements}
We are grateful to the anonymous referee for their comments which have helped to improve the quality of the manuscript.
N.J.M. acknowledges support from the Swedish National Space Agency (SNSA/Rymdstyrelsen).
P.F.L.M. acknowledges support from the following grants: ST/Y002563/1, UKRI1193.
A.H. acknowledges support from the Swedish Research Council.
We thank J. Southworth for helpful comments on the manuscript.
This research has made use of the SIMBAD database, operated at CDS, Strasbourg, France \citep{2000A&AS..143....9W}. 
This work has made use of data from the European Space Agency (ESA) mission \gaia{}\footnote{https://www.cosmos.esa.int/gaia}, processed by the \gaia{} Data Processing and Analysis Consortium DPAC\footnote{https://www.cosmos.esa.int/web/gaia/dpac/consortium}.
We acknowledge the use of public data from the \swift{} data archive.
Based on observations collected at the European Southern Observatory under ESO programmes 082.D-0499, 083.D-0549, 085.C-0614, 085.D-0395, 086.D-0078, 190.D-0237 and 112.25DR.
This research made use of Lightkurve, a Python package for \textit{Kepler} and \textit{TESS} data analysis \citep{2018ascl.soft12013L}.
This research made use of Astropy,\footnote{http://www.astropy.org} a community-developed core Python package for Astronomy \citep{astropy:2018}.
This research also made use of the HEASoft package \citep{2014ascl.soft08004N}.

\section*{Data Availability}

\tess{} light curves are available from the Mikulski Archive for Space Telescopes (MAST) -- https://archive.stsci.edu/. 
HARPS spectra are available from the ESO Science Archive Facility -- https://archive.eso.org/.
\swift{} UVOT data are available for download from NASA's High Energy Astrophysics Science Archive Research Center (HEASARC) -- https://heasarc.gsfc.nasa.gov/.


\bibliographystyle{mnras}
\bibliography{llaqr} 


\bsp	
\label{lastpage}
\end{document}